\documentclass[journals]{IEEEtran}

\usepackage[T1]{fontenc}
\usepackage{hyperref}
\usepackage[normalem]{ulem}
\usepackage{url}
\usepackage{graphicx}
\usepackage{adjustbox}
\usepackage{amssymb, amsmath}
\usepackage{amsfonts}
\usepackage{makecell} 
\usepackage{cite}
\usepackage{bm}
\usepackage{cases}
\usepackage[table, dvipsnames]{xcolor}
\usepackage{enumitem}
\usepackage{algpseudocode}
\usepackage{algcompatible}
\usepackage{algorithm}
\usepackage{caption}
\usepackage{booktabs}
\usepackage{multirow}
\usepackage{physics}
\usepackage{multicol}
\usepackage{float}
\usepackage{subcaption}
\usepackage{booktabs,dcolumn}
\floatstyle{ruled}
\usepackage{color,soul}
\newfloat{model}{thp}{lop}
\floatname{model}{Model}

\usepackage{multirow,mdframed}
\mdfsetup{skipabove=0pt,skipbelow=0pt}
\mdfdefinestyle{model}{%
  innertopmargin=0pt,
  innerbottommargin=0pt,
  innerleftmargin=0pt,
  innerrightmargin=0pt,
  skipbelow=0pt,%
  skipabove=0pt,
  splitbottomskip=0pt,
  splittopskip=0pt,
  leftmargin =0pt,%
    rightmargin=0pt,
  splittopskip=0pt,
    usetwoside=false,
}

\restylefloat{algorithm}

\definecolor{darkgreen}{RGB}{0,135,0}
\definecolor{coral}{RGB}{255,115,80}

\begin{document}
\title{Restoring Feasibility in Power Grid Optimization: A Counterfactual ML Approach}

\author{
Mostafa~Mohammadian,~\IEEEmembership{Student~Member,~IEEE,}
Anna Van Boven,~\IEEEmembership{Student~Member,~IEEE,}\\
        and~Kyri~Baker,~\IEEEmembership{Senior Member,~IEEE}
\vspace{-7mm}
\thanks{This research is supported by National Science Foundation awards $2041835$ and $2242930$.}
} 

\maketitle
\pagenumbering{gobble}

\begin{abstract}
Electric power grids are essential components of modern life, delivering reliable power to end-users while adhering to a multitude of engineering constraints and requirements.
In grid operations, the Optimal Power Flow problem plays a key role in determining cost-effective generator dispatch that satisfies load demands and operational limits. 
However, due to stressed operating conditions, volatile demand profiles, and increased generation from intermittent energy sources, this optimization problem may become infeasible, posing risks such as voltage instability and line overloads. This study proposes a learning framework that combines machine learning with counterfactual explanations to automatically diagnose and restore feasibility in the OPF problem.  Our method provides transparent and actionable insights by methodically identifying infeasible conditions and suggesting minimal demand response actions. 
We evaluate the proposed approach on IEEE $\bm{30}$-bus and $\bm{300}$-bus systems, demonstrating its capability to recover feasibility with high success rates and generating diverse corrective options, appropriate for real-time decision-making. These preliminary findings illustrate the potential of combining classical optimization with explainable AI  techniques to enhance grid reliability and resilience.
\end{abstract}


\IEEEpeerreviewmaketitle

\section{Introduction}
\IEEEPARstart{E}{lectric} power grids are essential components of modern society that ensure reliable electricity supply to end-users such as industries, businesses, and residential consumers \cite{Mostafa21}. At the heart of power grid operations lies the Optimal Power Flow (OPF) problem, which plays a crucial role in both planning and real-time decision-making. OPF seeks to determine the most cost-effective way to dispatch generators to meet fluctuating load demands while adhering to various physical and engineering constraints.  
OPF is further complicated by volatile renewable energy injections, rising electricity consumption, and an increasing necessity for tighter operating margins.
Moreover, since grid operators frequently have to operate systems close to their security limits to handle higher loading conditions, finding a feasible solution to OPF can become considerably more challenging~\cite{efs_operational}.


Previously, in power systems research, various techniques have been developed to better formulate and solve the large-scale OPF problem.  First, from an algorithmic perspective, researchers have developed a range of computationally efficient numerical methods to deal with larger and more complex networks \cite{donti2020dc3, 10328223, Mosi2024DynOPF}. Second, from a formulation viewpoint, OPF has broadened to include multiple objectives (e.g., cost minimization, emission reduction) as well as a range of physical and operational constraints \cite{10807052, MOHAMMADIAN2023109551}. 
Despite these advancements, solving the OPF problem is still challenging, and it can be nontrivial to find solutions that simultaneously satisfy all of the constraints. These limitations underscore the need for robust frameworks that can systematically diagnose, mitigate, and ultimately prevent infeasibilities within OPF domain \cite{GUNDA201623}.

When the OPF problem becomes infeasible — indicating that no solution meets the constraints — system operators face significant challenges to ensure reliability. Diagnosing the underlying causes of these infeasibilities is critical, yet traditional methods offer limited insights into adjustments that can resolve the infeasibilities.
In practice, operators often resort to manual interventions or time-consuming ``feasibility fixing'' procedures within optimization solvers. 
Under heavy load scenarios (where system constraints are highly binding), for instance, operators may reduce congestion on transmission lines by incorporating dynamic line ratings, or by temporarily relaxing transmission line constraints \cite{constraint_relax}. Alternatively, operators may adjust power injections to restore feasibility and prevent widespread outages, which necessitates a time-intensive post-processing procedure.

As mentioned above, most traditional solvers will take a long time to recognize and mitigate an infeasibility in a larger power system. However, resolving an infeasibility must happen rapidly in order to prevent voltage instability and grid collapse. Several papers have studied techniques for quickly identifying and localizing infeasible power grid conditions. In \cite{evaluate_feasibility, infeasible_dg}, an Equivalent Circuit Formulation (ECF) is utilized to determine how much additional current is needed in each bus to maintain power balance. 
Other works, such as\cite{els, gcnn} implement machine learning (ML) strategies that perform emergency load shedding in response to a contingency event, such as a line outage. 
These techniques are faster than traditional optimization solvers in identifying and resolving a power system infeasibility.
However, they are not fully prescriptive; none of these works specify both the exact amount and the location of the load to shed to maintain the feasibility of the system. 

Additional works incorporate explainable Artificial Intelligence (XAI) techniques into power systems operations \cite{MACHLEV2022100169}, partially to assist in understanding system behavior. 
 \cite{xai_uvls} applies XAI techniques to identify the causes of contingency events on the grid, demonstrating how XAI can  increase understanding of system behaviors and attributes that contribute to fault conditions.
In this context, an XAI technique called  counterfactual explanations \cite{Mothilal2019ExplainingML, brughmans2022nice} offers new ways to investigate power system operations by explaining the impact of small changes to system characteristics. 
In recent work, counterfactual explanations have been used to understand generator dispatch decisions in DC-OPF \cite{dc_cf}.  This type of insight, when integrated with robust, data-driven methods,  can enable system operators to identify, explain, and mitigate infeasible circumstances before they become costly or create unsafe conditions. 
Building on recent advances in machine learning and counterfactual explanations within power systems, our paper introduces an end‐to‐end learning framework designed to restore feasibility in OPF problems. As an initial foray into this area, we focus on DC Optimal Power Flow (DC-OPF) problems to understand if the use of counterfactual explanations in these problems can be useful.
Our integrated system first detects infeasible scenarios and then utilizes counterfactual explanations to pinpoint possible bus power injection adjustments needed to restore feasibility. This allows grid operators to have a suite of diverse options to restore feasibility. In contrast with the concept of ``interpretability'', which addresses transparency of the model itself, explainability seeks to explain the \emph{behavior} of a model - key for understanding OPF feasibility adjustments.
Together, this study aims to bridge the gap between classical optimization methods and modern AI-driven explainability, marking a significant step forward in the design of robust and interpretable power grid optimization tools \cite{MACHLEV2022100169}.

\section{Preliminaries}\label{sec: preliminaries}
This section presents the foundational concepts for DC-OPF and outlines the analytical feasibility restoration scheme that serves as our Baseline approach.
\subsection{DC Optimal Power Flow}\label{subsec: DC-OPF}
DC-OPF is widely used for current power system operations due to its convexity and computational benefits. In the DC-OPF framework, the aim is to determine the minimum-cost generator dispatch that meets load requirements within a power network, subject to power flow constraints on transmission lines and generator operating limits. 
Let us consider a power network with \(N \) buses, represented by the set \(\mathcal{N}\), and \(l\) transmission lines, forming the edge set \(\mathcal{E}\). 
 For each bus $i \in \mathcal{N}$, \(p^g_{i}\) and \(p^d_{i}\) denote the total generated power and power consumption (demand), respectively.  The variable \(p^f_{ij}\) indicates the active power flow on each transmission line, with $b_{ij}$ representing the line susceptance. In addition, the vector \(\boldsymbol{\theta} \in \mathbb{R}^n\) captures the phase angles at all buses, with \(\theta_{0}\) denoting the reference bus angle. Hence, the DC-OPF problem can be formulated as \cite{9810638}:

\begin{equation}\label{eq: cost_func}
    \underset{\boldsymbol{\theta}, \boldsymbol{p^g}}{\text{minimize:}} \quad \text{cost}(\boldsymbol{p^g}) = \sum_{i \in \mathcal{N}} c_{i}(p^g_{i}) 
\end{equation}
\qquad    subject to
\begin{align}
        &p^g_{i} - p^d_{i}  = \sum_{(ij) \in \mathcal{E}} p^f_{ij}   \hfill &i \in \mathcal{N} \label{eq: power_balance}\\
    &\underline{p}^g_{i} \leq p^g_{i} \leq \overline{p}^g_{i}   \hfill &i \in \mathcal{N}\label{eq: minmax_pow}\\
    & p^f_{ij} \leq \overline{p}^f_{ij} \hfill &(ij) \in \mathcal{E}\\
    & p^f_{ij} = b_{ij} (\theta_{i} - \theta_{j}) \hfill &(ij) \in \mathcal{E}\\
    & \theta_{0} = 0^{\circ} \hfill & i \in \mathcal{N} \label{eq: ref_angle}
\end{align}
where \(\underline{p}^g_i\) and \(\overline{p}^g_i\) denote the lower and upper limits on active power generation for each generator. The objective function \eqref{eq: cost_func}, which captures the cost of the generator response, is modeled as a quadratic function. The coefficient $c_i$ in the cost function defines the operational cost associated with the generator $g$. Equation \eqref{eq: power_balance} enforces the power balance in the bus \(i \in \mathcal{N}\). Meanwhile, constraints \eqref{eq: minmax_pow} through \eqref{eq: ref_angle} ensure that the generator output and the transmission line flows \(p^f_{ij}\) remain within acceptable operating boundaries.
\subsection{Baseline Feasible Solution Generation}\label{subsec: load_shed}
 
To provide a Baseline for comparison with our counterfactual-based feasibility restoration scheme, we formulate a convex problem that always guarantees feasbility with minimal load shed.
Here, power injections at each bus $i \in \mathcal{N}$ are adjusted by $p^{d_{\text{shed}}}$ to ensure feasibility:


\begin{equation}\label{eq: objective function}
    \underset{\boldsymbol{\theta}, \boldsymbol{p^g}, \boldsymbol{p^{d_{\text{shed}}}}}{\text{minimize:}} \quad  \big|\big| \bigl(p^d - p^{d_{\text{shed}}}\bigr) \big|\big|_k
\end{equation}
subject to the following constraints:
\begin{align}
    & p^g_{i} - p^{d_{\text{shed}}}_i = \sum_{(ij) \in \mathcal{E}} p^f_{ij}, 
        && i \in \mathcal{N}, \label{eq: power_balance_shed}\\
    & 0 \;\leq\; p^{d_{\text{shed}}}_i \;\leq\; p^d_{i}, 
        && i \in \mathcal{N}, \label{eq: constrained load_shed}\\
    & \underline{p}^g_{i} \leq p^g_{i} \leq \overline{p}^g_{i}, 
        && i \in \mathcal{N}, \label{eq: minmax_pow_shed}\\
    & p^f_{ij} \;\leq\; \overline{p}^f_{ij}, 
        && (ij) \in \mathcal{E}, \\
    & p^f_{ij} = b_{ij} \bigl(\theta_{i} - \theta_{j}\bigr), 
        && (ij) \in \mathcal{E}, \\
    & \theta_{0} = 0, 
        && i \in \mathcal{N}.\label{eq: ref_angle_shed}
\end{align}
where $k$ represents either the $L^1$ or $L^2$ norm, \(p^d_i\) represents the original load demand, and \(p^{d_{\text{shed}}}_i\) indicates the adjusted load demand at bus \(i\) after curtailments to ensure feasibility of the solution. By imposing the constraint \(0 \leq p^{d_{\text{shed}}}_i \leq p^d_{i}\), we ensure that the power injection adjustment process is physically viable. However, we can relax this constraint further to bring back the solution to a feasible region by introducing a range of other practical actions, such as bringing backup generation online at a specific bus, fictitious demand, or discharging a battery. The remaining constraints mirror those in the DC-OPF formulation. 

\section{Problem Formulation}\label{sec: prob_formulation}

Recently, counterfactual explanations have emerged as a strategy to
 provide ``\textit{what-if}'' scenarios by suggesting minimal modifications to input features that alter a model’s prediction. 
Specifically, the following objective function is proposed in \cite{Wachter2017CounterfactualEW} to generate a counterfactual $\bm{c}$ that achieves a target prediction $y$ with minimal changes to an input \(\bm{x} \in \mathbb{R}^d\): 
\begin{align}\label{eq: feasibility}
   \bm{c} = \arg\min_{\bm{c}} \Bigl(\text{yloss}(f(\bm{c}), y) + \lvert \bm{x} - \bm{c} \rvert \Bigr)
\end{align} 
where the first term, \(\text{yloss}(f(\bm{c}), y)\), nudges the counterfactual towards a new prediction, and the second term, \(\lvert \bm{x} - \bm{c} \rvert\), ensures that \(\bm{c}\) remains close to the original instance \(\bm{x}\). There are both model-agnostic and model-specific approaches for generating counterfactual explanations. This study concentrates on model-agnostic methods, which rely solely on inputs and outputs and do not require access to the internal structure of specific models.

Our counterfactual model takes as input a vector $\bm{x}$  that is an infeasible load profile, as well as a machine learning model $f: \mathbb{R}^d \xrightarrow{} \mathbb{R}$ that is trained to classify a load profile as feasible or infeasible. The counterfactual model then produces a set of $k$ counterfactual examples \(\{\bm{c}_1, \bm{c}_2, \dots, \bm{c}_k\}\), each of which is similar to the input $\bm{x}$, but perturbed enough so that the model $f$ classifies the counterfactual as feasible. 
The set of counterfactuals are \(d\)-dimensional as \(\bm{x}\), and we assume that \(f\) is static (i.e., it does not evolve over time). In the proposed framework for restoring feasibility, we aim to create a collection of counterfactuals that are not only realistic in relation to the original input (which is the load profile in our study) but also actionable, allowing users to implement the suggested changes effectively. 
It is important to note that counterfactual generation is a post-hoc process, which improves interpretability after the model has been trained.

Based on the method presented in \cite{Mothilal2019ExplainingML}, a unified objective function that optimizes over a set of \(k\) counterfactuals is considered as follows:
\begin{equation}\label{eq: CF_obj}
\begin{aligned}
C(\bm{x}) = \arg\min_{\bm{c}_1,\dots,\bm{c}_k} \Bigl(
    &\frac{1}{k}\sum_{i=1}^k \text{yloss}(f(\bm{c}_i), y) \\
    +\, &\frac{\lambda_1}{k}\sum_{i=1}^k \text{dist}\bigl(\bm{c}_i, \bm{x}\bigr) \\
    -\,& \lambda_2\,\text{dpp\_diversity}\bigl(\bm{c}_1,\dots,\bm{c}_k\bigr)
\Bigr).
\end{aligned}
\end{equation}
where the function \(\text{yloss}\bigl(f(\bm{c}_i), y\bigr)\) measures how effectively each counterfactual \(\bm{c}_i\) is pushed toward the desired outcome \(y\). The term \(\text{dist}\bigl(\bm{c}_i, \bm{x}\bigr)\) ensures that each counterfactual stays close to the original instance \(\bm{x}\), which reflects \textit{proximity} and supports actionability in real-world applications,  as shown in \eqref{eq: feasibility}. 
The dpp\_diversity term encourages variety among the $k$ counterfactual samples via a similarity kernel matrix. A higher value for dpp\_diversity indicates a more diverse set of CF explanations, so that the model doesn't predict $k$ slight perturbations of a single counterfactual. More information on the implementation of the diversity term can be found in \cite{Kulesza2012DeterminantalPP}.
Although having diverse CF examples may increase the chances that at least one example will be actionable for the user, examples may end up changing a large set of features or maximizing diversity by considering big changes from the original input. This issue could be worsened in high-dimensional feature spaces. That is why we need a combination of diversity and feasibility, as formulated in \eqref{eq: CF_obj}. Moreover, \(\lambda_1\) and \(\lambda_2\) are hyperparameters that control the relative importance of the three components within the overall objective function.



In addition to these three terms, a notion of sparsity is also considered—how many features need to be altered to transition \(\bm{x}\) into the counterfactual class. Because this sparsity constraint is non-convex, it is not included directly in the main objective; instead, it is addressed through a post-processing step on the generated counterfactuals \cite{Mothilal2019ExplainingML}. This multi-component formulation effectively balances the need to push each counterfactual toward the desired model output, keeping it sufficiently close to the original instance and ensuring diversity across the set of possible solutions.

Ultimately, one of the important practical considerations is the choice of the yloss function. A straightforward option for \(\text{yloss}\) could be either \(L^1\)- or \(L^2\)-loss. 
These loss functions focus on minimizing the distance of \(f(\bm{c})\) from the target \(y\); however, a valid counterfactual only needs \(f(\bm{c})\) to exceed or fall below the threshold set by \(f\) (usually $0.5$), rather than being as close as possible to the target \(y\) ($1$ or $0$). In fact, pushing \(f(\bm{c})\) to be near $0$ or $1$ often leads to significant alterations in \(\bm{x}\) towards the counterfactual class, making the resulting counterfactual less practical for operators. We just want the prediction to cross the boundary (e.g., flip from infeasible to feasible), not be extremely confident.
To address this, we adopt a hinge-loss function that imposes no penalty as long as \(f(\bm{c})\) remains above a certain threshold above $0.5$ when the desired class is $1$ (and below a specific threshold when the desired class is $0$). It applies a penalty that is proportional to the difference between \(f(\bm{c})\) and $0.5$ when the classifier is correct (but still within the threshold), and a more substantial penalty when \(f(\bm{c})\) fails to indicate the desired counterfactual class:
\begin{align*}
   yloss(f(\bm{c})) = \max\bigl(0,\; 1 - z \cdot logit\bigl(f(\bm{c})\bigr)\bigr), 
\end{align*}
where \(z = -1\) if \(y = 0\) and \(z = 1\) if \(y = 1\).
If the prediction already crosses the boundary confidently ($logit\bigl(f(\bm{c})\bigr) > 0$ for target class $1$), then $yloss = 0$.
Here, \(logit\bigl(f(\bm{c})\bigr)\) represents the unscaled output from the machine learning model (for instance, the final logits that are fed into a softmax layer for predictions in a neural network).
 It is important to note that while the optimization problem outlined in \ref{eq: CF_obj} is non-convex (due to the diversity term as well as the yloss), the model-agnostic search methods use heuristics to generate the counterfactual samples for non-differentiable models  like decision trees. These heuristics allow us to generate diverse and feasible counterfactuals efficiently without the need for gradient-based solvers or surrogate models. While this approach may sacrifice global optimality, it offers scalability and practical utility in producing actionable explanations for operators in near-real-time.

\begin{figure}[t!] 
\centering \vspace{1mm}
\includegraphics[scale = 0.9, width=\columnwidth, trim= {0 1.6cm 0 0.06cm}, clip]{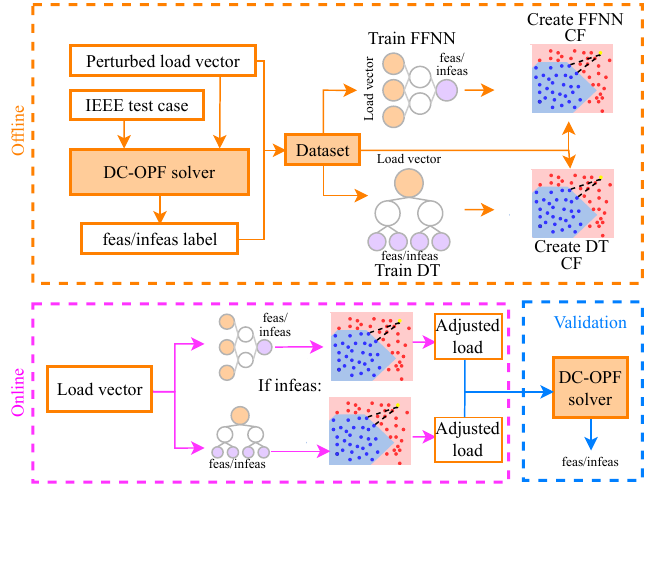} 
\caption{Proposed framework for infeasibility detection and resolution.} \label{fig:flow_diagram}\vspace{-5mm} 
\end{figure}

The illustration of the proposed framework is shown in Fig. \ref{fig:flow_diagram}.
The upper portion (orange) refers to the dataset creation by applying the DC-OPF solver to various load vectors, and the training of our two classifier models. The middle section (magenta) describes how we classify load vectors in real-time. If a load vector is found to be infeasible, our counterfactual mechanism adjusts the vector to feasibility. Finally, in the bottom right portion (blue), we validate the counterfactual-adjusted vector using a conventional solver to benchmark our counterfactual solutions and confirm feasibility.

\vspace{-3mm}
\section{Experimental Results} \label{sec: experimental result}
The effectiveness of the proposed method is evaluated on two power
networks of different sizes and complexity: IEEE $30$-bus and IEEE $300$-bus systems.  We also examine the flexibility of our counterfactual method by integrating two different classifiers—a traditional decision tree (DT) and a deep neural network (FFNN).
To benchmark our results, we compare them with the analytical solution of the Baseline presented in Section \ref{subsec: load_shed}, solved with Gurobi. 
The following sections provide details on the experimental setup and the results obtained from these two networks.

\paragraph{Dataset creation and training setting}
Topology, initial load data, and line limits for the IEEE $30$-bus and $300$-bus systems were obtained from the Power Grid Library (PGLib) repository \cite{PGlib-OPF}, including their respective network characteristics. In the $300$-bus system, line limits were scaled by a factor of $1.12$ to increase the likelihood of finding feasible solutions during the dataset generation.
We trained the two classification models—a feedforward neural network (FFNN) and DT—on a dataset \(\mathcal{D} = \{\left(\bm{p}^{d,i}, {y}^{i}\right)\}_{i=1}^{10,000}\). Each input \(\bm{p}^d\) represents a load vector, while its corresponding label \({y} \in \{0,1\}\) indicates feasibility in the DC-OPF setting, where $0$ indicates infeasibility and $1$ indicates feasibility of the DC-OPF problem.  To ensure robust model training, the dataset was balanced to include an equal number of feasible and infeasible instances, resulting in a $50/50$ split between the two classes. Following \cite{Mosi2024DynOPF}, each load demand \(p_i^d\) is generated by uniformly perturbing the nominal load \(p_i^d\), \(i \in \mathcal{N}\), by up to \(\pm 65\%\) in both test systems. Additionally, we only include samples where the total load is less than the total maximum generation and the load at each bus is less than the total maximum line flow into that bus.  In sum, every training sample represents a snapshot of the DC-OPF problem, consisting of a load profile and its feasibility indicator. The DC-OPF models are implemented and solved using Gurobi in Python. 
The dataset \(\mathcal{D}\) is divided into training and test sets in a $80/20$ split.


The FFNN and DT classification models were implemented in Python $3.9$ using PyTorch and scikit-learn on a personal Apple laptop with an M1 chip. The Adam optimizer was used to train the FFNN model across $300$ epochs at its default learning rate of \(10^{-3}\). 
The architecture of the FFNN included four hidden layers, each with $20$ neurons and ReLU activations, while the output layer utilized a softmax activation function. The decision tree was configured with a maximum depth of $4$.  
Since misclassifying samples as feasible is more costly and can lead to severe operational consequences, we incorporated a loss function that imposes a higher penalty on misclassified infeasible samples than on feasible samples.
After training, we assessed the performance of each classifier on the test dataset, and the accuracy results are presented in Table \ref{tab: Accuracy classifier}. These trained models form the foundation for generating counterfactuals in our framework. In this implementation, the counterfactual hyperparameters are fixed at \(\lambda_1 = 0.5\) and \(\lambda_2 = 1\). These values were determined through a grid search that balanced diversity and proximity.

\begin{table}[t!]
\centering \vspace{2mm}
\caption{Classification accuracy of the trained models on the test datasets for the IEEE $30$-bus and $300$-bus systems.} 
\label{tab: Accuracy classifier}
\begin{tabular}{l c c}
\toprule
Method & IEEE $30$-bus & IEEE $300$-bus \\ \midrule
FFNN &  $99.35\%$ & $86.30\%$ \\
DT  & $99.00\%$  &  $92.00\%$ \\
\bottomrule
\end{tabular}\vspace{-4mm}
\end{table}

\vspace{-3mm}
\subsection{Restoring Feasibility of the OPF Solution}
To evaluate our framework’s capability to restore feasibility for OPF scenarios that were initially classified as infeasible, we compare the approaches: 
counterfactuals generated via our feedforward neural network ($\text{CF}_{FFNN}$) or decision tree ($\text{CF}_{DT}$), and the Baseline (\ref{subsec: load_shed}).
The reported values represent the percentage of infeasible test instances that were successfully “fixed” and returned to a feasible state.

In both the IEEE $30$-bus system and the $300$-bus system, \(\text{CF}_{FFNN}\) and \(\text{CF}_{DT}\) each successfully restore every infeasible test case to a feasible state (as determined by the DC-OPF formulation described in Section \ref{subsec: DC-OPF}), achieving a $100\%$ feasibility recovery rate. This consistent performance demonstrates the robustness and adaptability of our counterfactual-based framework, even when applied to larger and more complex networks. 
Furthermore, these results show that both classifiers, when combined with the counterfactual generation method,  achieve high success rates that are comparable to commercial-grade solvers.

\begin{figure}[t!] 
\centering 
\includegraphics[scale= 0.83, width=\columnwidth]{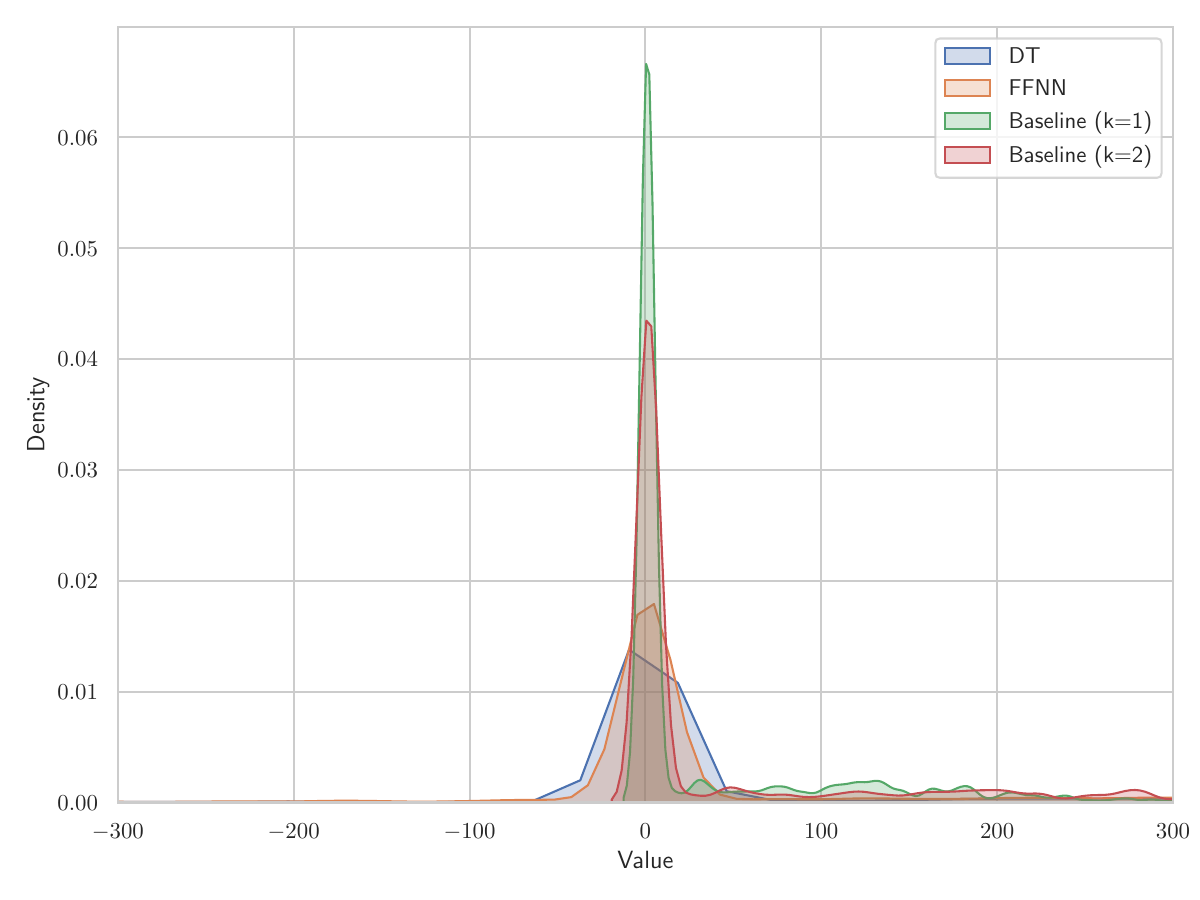}\vspace{-2mm} 
\caption{Distribution of overall bus-level power adjustments (perturbations) for the IEEE $300$-bus system.} \vspace{-4mm}\label{fig:load_shedding_dist} 
\end{figure}

In addition to the high feasibility restoration rates, we also examine the \textit{extent} of load adjustments (or ``perturbations'') needed by each method. 
Table \ref{tab: feasibility_restoring} summarizes the  mean net power injection adjustment
values over all the buses (along with standard deviations) for the IEEE $30$-bus and IEEE $300$-bus systems. 
The Baseline analytical solution achieves the guaranteed lowest average adjustments. Although our counterfactual-based (\(\text{CF}_{FFNN}\) and \(\text{CF}_{DT}\)) solutions have larger adjustments in the load, with a wider distribution, they remain competitive and their required perturbations often align closely with the Baseline.
Figure \ref{fig:load_shedding_dist} further illustrates these distributions for the $300$-bus case.
As illustrated, negative values correspond to bringing additional generation online or increasing load at specific buses, whereas positive values indicate the extent of load reduction in our framework.
While the Baseline solution achieves slightly smaller load adjustments, the two counterfactual methods follow a similar overall shape and frequency of perturbation.
The sharp peak around zero reflects that many bus-level perturbations are minimal or zero.
This is consistent with the goal of generating minimal corrective changes. Only infeasible samples are used in this analysis; the perturbation is computed between each infeasible sample and its corresponding feasible corrected version.

A key advantage of our approach is its flexibility in applying user-defined constraints or bounds on the adjustments—beyond standard load-shedding or demand-response strategies. For instance, operators can incorporate specific demand response mechanisms in certain buses, which may be preferable in systems where load curtailment must be carefully targeted or distributed for equity reasons. 
As a result, while our method may require a  higher mean power adjustment than the true optimal solution, it can still produce operationally appealing solutions that reflect real-world priorities (e.g., fairness \cite{sundar2024e}, local reliability criteria) without losing much in terms of total reduction. Overall, these findings highlight the practicality of counterfactual generation as an alternative or complementary method for restoring feasibility, offering both robust performance and considerable adaptability to operator-defined constraints.


\begin{table}[t!] \vspace{2mm}
\centering
\caption{Statistics of the load adjustments (in MW) required to restore feasibility.} 
\label{tab: feasibility_restoring}
\begin{tabular}{l c c | c c}
\toprule
\multirow{2}{*}{Method} & \multicolumn{2}{c|}{IEEE 30-bus} & \multicolumn{2}{c}{IEEE 300-bus} \\ 
\cmidrule{2-3} \cmidrule{4-5}
 & Mean  & Std.  & Mean  & Std.\\
\midrule
$\text{CF}_{FFNN}$  & $11.03$  & $6.76$  & $88.62$  & $236.88$\\
$\text{CF}_{DT}$  & $11.14$  & $6.84$  & $93.34$  & $298.26$\\
Baseline  & $5.33$  & $3.18$  & $53.75$  & $102.70$\\
\bottomrule
\end{tabular}\vspace{-3mm}
\end{table}




\subsection{Diversity in Generated Counterfactuals}  
Another advantage of our counterfactual-based approach is the ability to produce multiple feasible solutions (\(k\) options) for the same infeasible scenario, thereby providing flexibility for operators or planners. In the IEEE $30$-bus system, for instance, there are $20$ buses that contain loads, and restoring feasibility with minimal changes typically requires adjustments to the same subset of buses. Our approach generates feasibile solutions that vary both the buses with adjusted load and the amount of load adjustment at each bus. 

Table \ref{tab:CF_diversity} compares three distinct counterfactual options derived from both the DT (CF\(_\text{DT}\)) and FFNN (CF\(_\text{FFNN}\)) classifiers,  showing how these counterfactuals can restore feasibility under different load profiles. For one of the infeasible load profiles in the system, in CF\(_\text{DT}\) framework, Option \#1 necessitates moderate load reductions at Bus $8$ ($16.903$ MW) and Bus $10$ ($1.869$ MW), while Option \#2 primarily reduces load  at Bus $8$ ($12.918$ MW) with smaller adjustments at other locations. Option \#3 involves a larger curtailment at Bus $8$ ($28.383$ MW) but also targets Bus $16$ ($2.947$ MW).
A similar trend is seen in CF\(_\text{FFNN}\), where Option \#1 reduces Bus $8$ ($3.8$ MW) and Bus $20$ ($0.6$ MW), but Option \#2 results in a larger reduction at Bus $8$ ($29.4$ MW) with a smaller adjustment at Bus $12$ ($0.8$ MW). Finally, Option \#3 balances the curtailment across Bus $8$ ($2.8$ MW) and Bus $16$ ($3.8$ MW).
While only three solutions are presented here, the proposed framework can produce 10 or more unique counterfactuals for the same infeasible scenario, enhancing grid operations.


\begin{table*}[t!]
\vspace{2mm}
\caption{Example counterfactual solutions showing different load reductions to restore feasibility.}
\label{tab:CF_diversity} \vspace{-1mm}
\centering
\begin{subtable}[t]{\textwidth}\centering
\begin{tabular}{l c c c c c c c  c c c c c c c c c c}
\toprule
Method & \multicolumn{16}{c}{Bus number} \\
\cmidrule(lr{0.5em}){2-18}
$\text{CF}_{DT}$ & 5& 6& 7& 8& 9& 10& 11& 12& 13& 14& 15& 16& 17& 18& 19& 20& 21\\
\midrule
\rowcolor{lightgray}
Option \#1 &  0& 0& 0& 16.903& 0& 1.869& 0& 0& 0& 0& 0& 0& 0& 0& 0& 0& 0 \\
Option \#2 &  0& 0& 0& 12.918& 0& 0& 0& 0& 0& 0& 0& 0& 0& 0& 0& 0.257& 0 \\
\rowcolor{lightgray}
Option \#3 &  0& 0& 0& 28.383& 0& 0& 0& 0& 0& 0& 0& 2.947& 0& 0& 0& 0& 0 \\
\hline
\end{tabular}
\end{subtable}

\begin{subtable}[t]{\textwidth}\centering
\begin{tabular}{l c c c c c c c  c c c c c c c c c c}
\toprule
Method & \multicolumn{16}{c}{Bus number} \\
\cmidrule(lr{0.5em}){2-18}
$\text{CF}_{FFNN}$ & 5& 6& 7& 8& 9& 10& 11& 12& 13& 14& 15& 16& 17& 18& 19& 20& 21\\
\midrule
\rowcolor{lightgray}
Option \#1 &  0& 0& 0& 3.775& 0& 0& 0& 0& 0& 0& 0& 0& 0& 0& 0& 0.555& 0 \\
Option \#2 &  0& 0& 0& 29.437& 0& 0& 0& 0.807& 0& 0& 0& 0& 0& 0& 0& 0& 0 \\
\rowcolor{lightgray}
Option \#3 &  0& 0& 0& 2.801& 0& 0& 0& 0& 0& 0& 0& 3.777& 0& 0& 0& 0& 0 \\
\hline
\end{tabular}
\end{subtable}\vspace{-2mm}
\end{table*}

In practice, such diversity can be critical: bus level power adjustment which sometimes require load-shedding, is not always granular or equitable, and system operators often face sociopolitical constraints—rotating outages, for example, might be preferable to consistently curtailing the same community. By offering three unique counterfactual “fixes,” our method allows an operator to weigh these trade-offs. They might opt to minimize total load reduction, pursue a more equitable distribution across multiple buses, or target specific industrial loads capable of demand response.  

While the Baseline feasibility restoration provides a single optimal solution, it does not offer multiple alternatives by default. Our counterfactual framework, in contrast, supplies a menu of feasible options, each meeting the same ultimate goal of restoring system feasibility but doing so through different operational adjustments. This increased flexibility is especially valuable in real-time or near real-time settings, where operators must balance engineering objectives, economic costs, and the practical realities of grid management.
\vspace{-4mm}
\subsection{Computation Time}
Table \ref{tab: computation time} presents the average computation times along with standard deviations for both the Baseline approach which is solved by Gurobi commercial solver and for our counterfactual-based methods (\(\text{CF}_{FFNN}\) and \(\text{CF}_{DT}\)), applied to both the $30$-bus and $300$-bus systems.  
As demonstrated by the results, both counterfactual approaches achieve significantly faster runtimes than the Baseline solution—up to a $43\times$ improvement for the CF\(_{DT}\) method on the IEEE $300$-bus system.
Furthermore, as the size of the network increases, the proposed methodology exhibits only a slight increase in computational time, showcasing its ability to manage larger systems efficiently.
It is worth mentioning that the Baseline optimization is a convex problem (\ref{subsec: load_shed}), which results in a significantly shorter computation time. This efficiency would deteriorate significantly when transitioning to nonlinear AC-OPF. On the other hand, the computation time for our counterfactual method remains constant regardless of whether the underlying problem is linear or nonlinear.

Lastly, because our counterfactual framework can generate multiple feasible solutions in a single pass, system operators have access to diverse remedial options without incurring any substantial additional time cost. This feature is not available in the mathematical optimization approach, as it always yields a single solution. In practice, this flexibility can be crucial when balancing economic, operational, and reliability considerations under tight scheduling constraints.

\begin{table}[t!]
\centering
\caption{Average computation times in (ms) (mean$\pm$standard deviation) for the considered systems.} 
\label{tab: computation time}
\begin{tabular}{l l l}
\toprule
Method & IEEE $30$-bus & IEEE $300$-bus \\ 
\midrule
$\text{CF}_{FFNN}$ &  $4.14  \pm 1.55$ & $41.0 \pm 7.23 $ \\ 
$\text{CF}_{DT}$  & $3.54  \pm 1.86$  &  $5.33  \pm 2.41 $ \\
Baseline            & $41.4  \pm 19.0$ & $234 \pm 15.1$\\
\bottomrule
\end{tabular}\vspace{-4mm}
\end{table}
\vspace{-2mm}
\section{Conclusion} \label{sec: conclusion}
We have introduced a framework that merges machine learning with counterfactual explanations to identify and address infeasibilities in DC-OPF solutions. Our approach effectively pinpoints the minimal changes in bus-level power adjustments  required to restore feasibility, as shown in both the IEEE $30$-bus and $300$-bus test systems. 
The experimental findings indicate that our method not only achieves high feasibility recovery rates identical to traditional optimization solvers but also provides various remedial options for grid operators, thus enhancing the interpretability and flexibility of the actions taken.
Our results are encouraging for this proof-of-concept framework. However, additional work is necessary to expand this research to larger, more complex networks, as well as to investigate its use in AC-OPF scenarios. In our future research, we will take into account additional corrective techniques, such as generator re-dispatch and demand response, to create a more comprehensive decision-support tool for system operators.
In general, this study connects classical optimization methods with data-driven techniques, presenting a practical and interpretable solution that can improve grid reliability under increasingly demanding operating conditions.

\vspace{-2mm}

\bibliography{references.bib}{}
\bibliographystyle{IEEEtran}

\end{document}